\documentclass[showpacs,aps,pre,floatfix]{revtex4}
\usepackage{graphicx} 

\begin{document}

%__________ Title, authors, institutions __________
\title{Revisiting Kawasaki dynamics in one dimension}

\author{M. D. Grynberg} 

\affiliation{Departamento de F\'{\i}sica, Universidad Nacional de  
 La Plata,(1900) La Plata, Argentina}

%________ Abstract ____________________
\begin{abstract}
Critical exponents of the Kawasaki dynamics in the Ising chain are re-examined
numerically through the spectrum gap of evolution operators constructed both
in spin and domain wall representations. At low temperature regimes the latter 
provides a rapid finite-size convergence to these exponents, which tend to 
$z \simeq 3.11$ for instant quenches under ferromagnetic couplings, while 
approaching to $z \simeq 2$ in the antiferro case. The spin representation
complements the evaluation of dynamic exponents at higher temperature 
scales, where the kinetics still remains slow.
\end{abstract}
%______________________________________

%________ Pacs Numbers _______________

\maketitle

%____________________________
\section{Introduction}
%____________________________

Kinetic Ising models have long been helpful in understanding general questions
of nonequilibrium statistical physics, as well as for elucidating particular issues 
of experimental relevance \cite{Bray}. There is by now a vast body of research
studying these models under Glauber and Kawasaki dynamics \cite{Glauber, 
Kawasaki}, the so called  models $A$ and $B$ respectively in the terminology of 
Hohenberg and Halperin \cite{Hohenberg}. The first type of dynamic considers 
single spin flip processes for describing relaxation towards equilibrium in a variety
of magnetic materials,  while the second case consists of spin exchanges specially 
aimed to study stochastic processes under conservation of total magnetization.
This constraint makes relaxation slower,  and is instrumental in studying phase
separation, domain growth, and freezing observed after rapid cooling in several
systems, such as binary fluids and alloys \cite{Bray}. All these phenomena are
characterized by a long relaxation time $\tau$ which near critical points diverges as
$\tau \propto \xi ^z$, where $\xi$ denotes the spatial correlation length (proportional
to the typical system size), and $z$ is the dynamic exponent of the universality class
to which the dynamic belongs.

When it comes to one dimension ($1d$), these dynamics are still largely amenable
to experimental probe \cite{Privman} apart from being interesting in their own
right \cite{Godreche}, and as is well known, in the Glauber case allow for an exact
treatment \cite{Glauber, Privman}. In contrast to that latter aspect, the equations
of motion for the Kawasaki dynamics conform an infinite hierarchy which to our
knowledge do not admit analytic solution, so the problem can only be approached
from approximations \cite{Privman}. Even though there is no finite critical
temperature through which to quench, using random walk arguments to derive an
approximate kinetic equation for the kink density under ferromagnetic (F)
interactions \cite {Robin}, interestingly,  the dynamic exponent has been put
forward to be the same as observed in higher dimensions, namely  $z = 3$, after a
deep quench to a low temperature (LT). Although for $d \ge 2$ such exponent was
also accounted for by  renormalization group arguments \cite{Bray, Bray2} as well
as by surface dynamical considerations \cite{Huse}, other studies in $1d$ related
systems \cite{Luscombe} suggest instead a value of $z=5$ \cite{CST}, in
agreement with linear response schemes \cite{Zwerger}.

In this work we re examine this exponent by constructing and diagonalizing 
numerically the kink evolution operator associated to the master equation of the
Kawasaki dynamics in finite chains. Following the physical picture given in 
\cite{Robin}, at late evolution stages  and LT regimes such operator would 
essentially describe a highly diluted system as the kink or domain wall density is 
basically a measure of the average inverse domain size. Consequently, in evaluating
dynamic exponents from finite samples it might be expected that size effects will 
pose no severe limitations for the kink representation.  In fact, as we shall see, the 
kink density will get fairly small even for the low lying levels of the evolution 
operator, that is what ultimately matters at large times. On the other hand there is 
also the question about estimating dynamic exponents at finite temperatures for 
which the  Kawasaki kinetics still remains critical,  as opposed to the exponential 
decay of the Glauber dynamics in $1d$. Although in this situation dynamic 
exponents are no longer related to domain growth - these are cut off by a finite 
correlation length - they still provide the fundamental relation between the typical 
system size and its relaxation time. We shall also address this issue by diagonalizing
the evolution operator of the original spins. Insofar temperature is not too low, these 
latter become weakly correlated and finite-size effects will not be paramount. 
A posteriori, our results will lend further support to this view.

Therefore these dual representations - kinks and spins - provide a means for probing
the robustness of dynamic exponents throughout different regimes. As a result, it
will turn out that for F couplings $z$ varies continuously from $\simeq 3.11$ to 
$\simeq 2$ at low and high temperature (HT) scales respectively. Such 
nonuniversality should come as no surprise, since in the HT limit the Kawasaki 
dynamics simply reduces to a diffusive disordering kinetics. However under 
antiferromagnetic (AF) exchanges these exponents are rather robust and, in line 
with previous findings \cite{Luscombe}, their values will remain diffusive even at
LT regimes, just as the $z$ of the Glauber universality class. We shall return to this
question later on within a context of antikink operators.

The layout of this work is organized as follows. In Sec. II we recast the 
$1d$-\,Kawasaki dynamics in terms of a quantum spin analogy that readily lends 
itself to evaluate spectrum gaps of evolution operators, i.e. dynamic exponents,
both in the spin and domain wall representations. Due to detailed balance, either 
of these descriptions can be brought to a symmetric representation by means of 
simple non-unitary spin rotations. This simplifies considerably the subsequent 
numerical analysis of Sec. III in which spectrum gaps are obtained via standard 
recursive techniques \cite{Lanczos} in various situations. For now let us remark 
that already modest chain lengths are able to yield clear finite-size trends both at 
LT and HT scales. Finally,  Sec. IV contains a  summarizing discussion along with
some remarks on extensions of this work.

%_____________________________________________________
\section{Spin and kink representations}
%_____________________________________________________

As is well known, the Ising model has no intrinsic dynamics because all spin 
operators involved in its Hamiltonian commute with one another. Therefore, an 
{\it ad hoc} dynamics must be prescribed by coupling the system to a heath bath 
at temperature $T$ so as to induce energy changes in the model. This is described 
by a master equation \cite{Kampen} for which we introduce briefly here some 
preliminary considerations. 

Basically, the dynamics is associated to a gain-loss equation, constructed 
generically as
\begin{equation}
\label{master}
\partial_t \,P(s,t) = \sum_{s'} \left[\: W (s'\to s)\, P(s',t)\,
- \, W (s \to s') \,P(s,t)\:\right]\,,
\end{equation}
which governs the evolution of the probability $P (s,t)$ that the system will be at 
state $\vert s \rangle$ at time $t$. The elementary change steps are embodied in 
the transition probability rates $W ( s \to s')$ per unit time at which configuration
$\vert s \rangle\,$ evolves  to $\vert s' \rangle\,$. For our purposes, it is convenient
to think of  this equation as a Schr\"odinger evolution in an imaginary time, namely 
$\partial_t \vert P (t)\, \rangle = - H \vert P (t) \,\rangle$, under a pseudo Hamiltonian
or evolution operator $H = H_d + H_{nd}$ whose diagonal and non-diagonal matrix 
elements are given by
\begin{eqnarray}
\label{diag}
\langle\,s\,\vert\, H_d\,\vert\,s\,\rangle &=& \sum_{s'\ne s}\, W(s \to s')\,,\\
\label{non-diag}
\langle\,s'\,\vert\,H_{nd}\,\vert\,s\,\rangle &=& -\,W(s \to s')\,.
\end{eqnarray}
Formally, this enables one to derive all subsequent probability distributions 
$\vert P(t) \,\rangle \equiv \sum_s P (s,t)\, \vert s \rangle\,$ from the action of $H$ 
on a given initial distribution, i.e. $\vert P(t) \,\rangle = e^{- H\,t} \vert P(0)\,
\rangle$ \cite{Kampen}. In particular, the relaxation time of any observable is 
singled out by the eigenvalue $\lambda_1$ corresponding to the first excitation 
mode of $H$, i.e. $\tau^{-1} = {\rm Re}\,\lambda_1 > 0$, whereas the steady 
state merely corresponds to an eigenvalue $\lambda_0 = 0$\, \cite{Kampen}. 
If the steady configuration should actually coincide with the Boltzmann equilibrium
distribution, the above rates must be constrained by detailed balance, that is 
$W (s \to s') \, e^{- \beta E (s)} =  W (s' \to s) \,e^{- \beta E (s')}$,  $\forall \,s,\,s',$
where $E$ stands for the respective energy configurations of the system in question,
and $\beta \equiv 1/k_B T$. 

Detailed balance itself can not determine entirely the form of such rates, thus for 
the specific case of the Kawasaki dynamics hereafter considered we take up the 
common choice \cite{Bray,Kawasaki} 
\begin{equation}
\label{rates0}
W (s \to s') = \frac{1}{2}\Big[\,1- \tanh \Big(\,\frac{\beta }{2} \,
\Delta E_{s, s'} \Big)\,\Big]\,,
\end{equation}  
where $\vert s \rangle ,\, \vert s'\rangle $ are states differing at most in a pair of 
nearest neighbor (NN) spins exchanged at some location $i = 1,2,\,...\,L$, and  
$\Delta E _{s , s'} \equiv E (s') -  E (s)$ is the change of the Ising energy $E = - J 
\sum_i s_i \,s_{i+1},\, (s_i = \pm 1)$ either with F $(J > 0)$, or AF $(J < 0)$ 
interactions. Therefore depending on the spin states at locations $i-1$ and $i+2$,
the rate at which spins $(s_i,\, s_{i+1}) = (s,\,-s)$ exchange their states results in
\begin{equation}
\label{rates}
R_{i-1,i+2} (\pm K) = \frac{1}{2} \pm \frac{\alpha_{_K}}{4} (\,s_{i+2} - s_{i-1}) 
\,,
\end{equation} 
where $K = \beta J$, $\alpha_{_K} = \tanh 2 K$, and the signs $\pm$ denote
forward and backward hoppings as depicted schematically in Fig.\,1.
%_______________ Fig. 1 (schematic view) _______________________________
\begin{figure}[htbp]
\vskip -8.5cm
\centering
\includegraphics[width=0.8\textwidth]{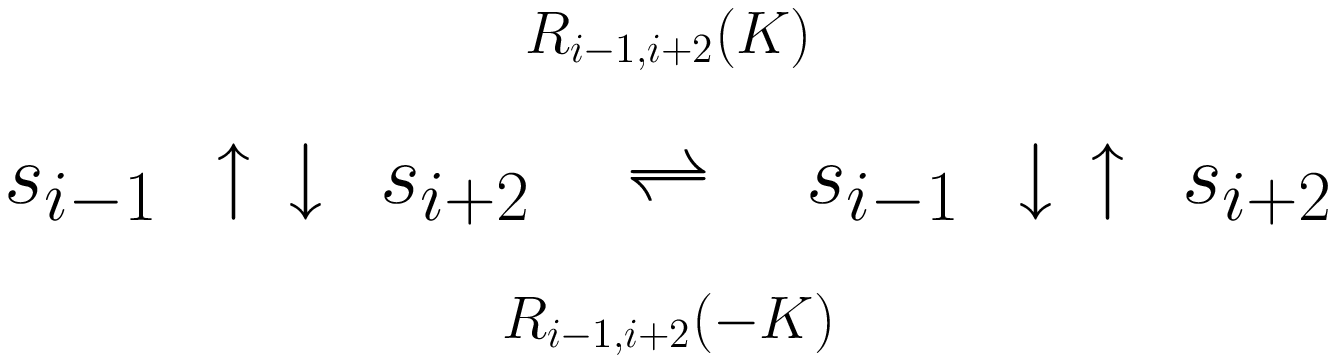}
\vskip -9cm
\caption{Transition probability rates for neighboring spin 
exchanges, as defined in Eq.\,(\ref{rates}).}
%\label{}
\end{figure}
%______________________________________________________________________
If we think of the spin configurations $\vert s \rangle$ as being already diagonal
in the $z$-direction, say, then by promoting $s_i$ variables to Pauli matrices 
$\sigma_i^z$, the operational analog of Eq.\,(\ref{non-diag}) will read
\begin{equation}
\label{nondiag-spins}
H_{nd} = - \sum_i  \left[\, R^z_{i-1,i+2} (K) \,\sigma^+_{i+1}\sigma^-_i + \,
R^z_{i-1,i+2} (-K)\, \sigma^+_i \sigma^-_{i+1}\right] \,,
\end{equation}
where $\sigma^+, \sigma^-$ are the usual spin-$\frac{1}{2}$ raising and
lowering operators, while $R^z_{i-1,i+2} (\pm K) = \frac{1}{2} \pm 
\frac{\alpha_{_K}}{4} (\,\sigma^z_{i+2} - \sigma^z_{i-1}) \,,$ is simply the 
operational counterpart of Eq.\,(\ref{rates}).  As for the diagonal elements of 
(\ref{diag}) needed for conservation of  probability, notice that these basically count
the number of hoppings in which a given configuration can evolve to different ones
by exchanging NN spins at a time.  This can be properly tracked down in terms of 
number operators $\hat n = \sigma^+ \sigma^- $ probing and weighting all NN 
spins, namely
\begin{equation}
\label{diag-spins}
H_d = \sum_i \left[\,  R^z_{i-1,i+2} (K) \,\hat n_i\, (1- \hat n_{i+1})
+ R^z_{i-1,i+2} (-K)\,\hat n_{i+1}\, (1- \hat n_i)\,\right]\,.
\end{equation}

Although the correlated exchange terms of Eq.\,(\ref{nondiag-spins}) leave us with
a non-symmetric evolution operator, in preparation for the numerical analysis of 
Sec. III  we can make some progress by exploiting detailed balance. This latter 
warrants the existence of representations in which $H$ is symmetric and thereby 
diagonalizable \cite{Kampen}. For our purposes, it suffices to consider the diagonal
non-unitary similarity transformation 
\begin{equation}
\label{trans1}
S = e^{-\frac{K}{2}\,\sum_j \sigma^z_j \, \sigma^z_{j+1} }\,,
\end{equation}
under which the hopping terms of $H_{nd}$ transform as
\begin{equation}
\label{transform}
\sigma^{\pm}_{i+1} \sigma^{\mp}_i \to \exp \left[\,\mp K \left( \sigma^z_{i+2}
- \sigma^z_{i-1 }\right)\, \right]\, \sigma^{\pm}_{i+1} \sigma^{\mp}_i\,,
\end{equation}
while leaving invariant all the above mentioned diagonal operators. Hence, after 
straightforward manipulations it can be readily verified that ${\rm H}_{spin} = 
S  H S^{-1}$ actually produces a self adjoint spin representation which for periodic 
boundary conditions (PBC) is found to be
\begin{eqnarray}
\nonumber
{\rm H}_{spin} &=& - \frac{1}{8} (\,1 + \gamma_{_K} ) \, \sum_i \, \left(\, 1 + 
\tanh^2 \! K \,\sigma^z_{i-1}\sigma^z_{i+2}\right)
 \left( \, \sigma^x_i \sigma^x_{i+1} + \sigma^y_i \sigma^y_{i+1}\,  \right)\\   
\label{hermitian-spin}
&+& \frac{1}{4} \sum_i \left[\,1 \,+\,\alpha_{ _K} \,\sigma^z_i \sigma^z_{i+2}
 \,-\, (1+ \alpha_{ _K}) \,\sigma^z_i \sigma^z_{i+1} \, \right]\,,
\end{eqnarray}
where $\gamma_{_K} = {\rm sech}\, 2 K$. 
In passing, it is worth pointing out that in the HT limit this operator coalesces into a
Heisenberg ferromagnet, i.e. ${\rm H}_{spin} \to -\frac{1}{4} \sum_i  \left(\,\vec 
\sigma_i \cdot \vec\sigma_{i+1} - 1  \right)$, with a spectrum gap behaving like 
$\propto 1/L^2$, as it should for a disordering diffusive kinetics. On the other hand,
in such regime the equilibrium correlation length becomes of the order of the lattice
spacing, thus the numerical diagonalization of ${\rm H}_{spin}$ in finite chains 
can be a reasonable starting point to evaluate dynamic exponents as temperature 
is quenched slightly (though {\it instantly} to a constant value, because transition 
rates have been taken time independent  throughout). 

When it comes to LT regimes however, Eq.\,(\ref{hermitian-spin}) is not of much 
practical use, since the  system approaches a state consisting of long magnetic 
domains whose typical sizes are of the order of the correlation length $\xi \sim 
e^{2 \vert K \vert}$ (nonetheless, for $J > 0$ see the upper bounds to $z$ and their
extrapolations provided in Sec. III A). In this more interesting scenario the dynamics
is strongly affected by whether the coupling exchanges are F or AF, thus it is
convenient at this point to separate the discussion accordingly.

%________________________________________
\subsection{Kink representation: $J > 0$}
%________________________________________

\begin{figure}[htbp]
\vskip -6.75cm
\centering
 \includegraphics[width=0.65\textwidth]{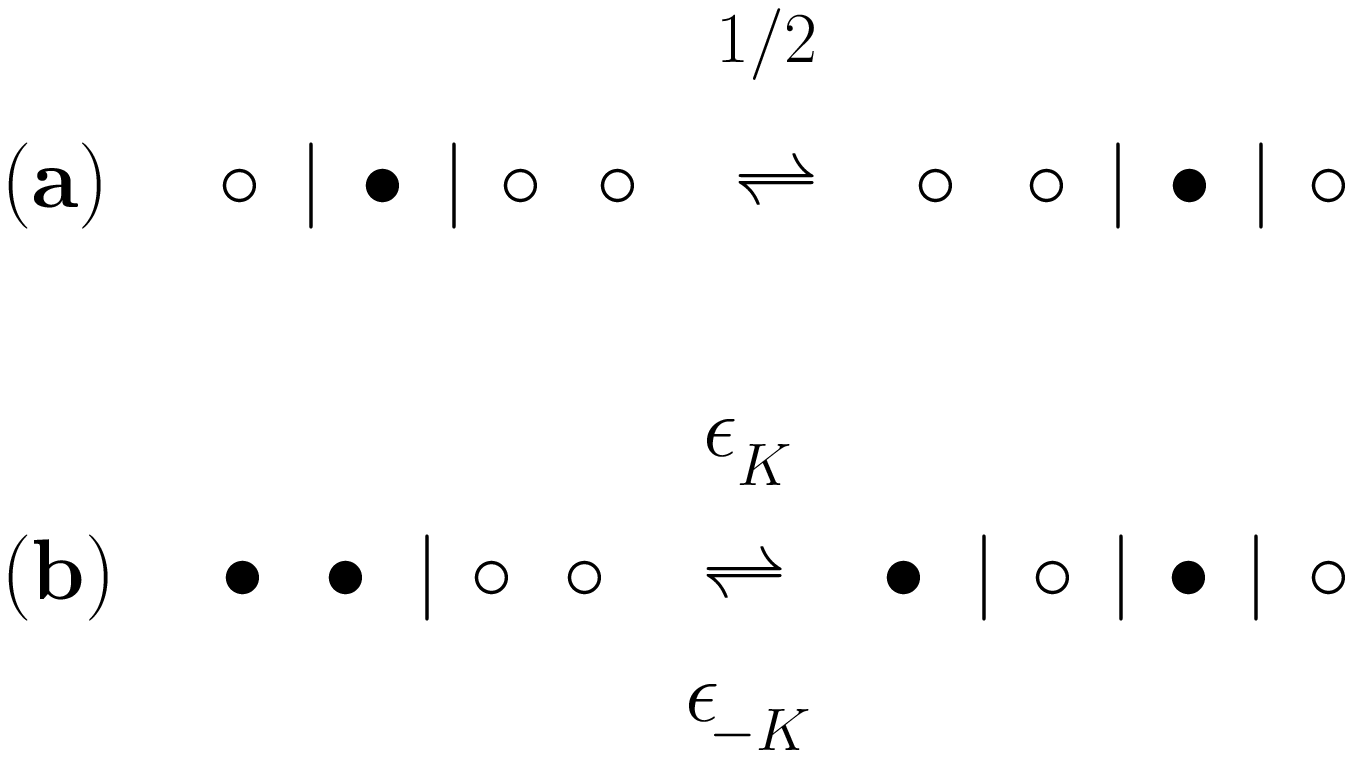}
\vskip -5.25cm
\caption{Transition rates for (a) diffusion of double kinks, and (b) assisted 
deposition-evaporation of kinks. These latter are denoted by vertical lines 
separating domains of opposite spin orientations.}
\label{Fkinks}
\end{figure}

As was mentioned in Ref.\,\cite{Robin}, the dynamics of this situation is basically 
mediated by single spins detaching from a domain wall (a kink),  and then 
performing a random walk. Some of  these spins will return to their original 
domains, while others will reach the next ones. On average, as a result of many 
of these processes the domains themselves will perform a random walk until they
meet and merge into larger domains. Besides, single meandering spins may
eventually nucleate and trigger the growth of a new domain. The key issue in this 
description is that for large times very few kinks will survive at LT scales.
In particular, in equilibrium these are totally uncorrelated and with an average 
density $\sim 1/\xi$, whereas that of the spins mediating the whole process is
of order $1/\xi^2$ as they are made up of two consecutive kinks (see 
Fig.\,\ref{Fkinks}a below).  Thus, for large times it makes sense to attempt a 
nonequilibrium description in terms of kinks rather than spins. Presumably 
(as confirmed later on in Sec. III), the first excitation mode of the kink evolution 
operator will also correspond to a diluted state, so size effects in a numerical 
estimation of $z$ will not be as severe as in the spin description.

%_______________ Fig. 2 (Kinks: schematic view) _____________________
%_______________________________________________________________

To recover the equilibrium behavior of the Ising model from now on we consider 
the case of zero magnetization, so this dual representation corresponds to 
the two-to-one mapping outlined schematically in Fig.\,\ref{Fkinks}. If we imagine
these kinks as hard core particles $A$, then the Kawasaki dynamics involve two 
basic processes (i) dimer diffusion $A+ A + \emptyset \rightleftharpoons 
\emptyset + A + A$, and (ii) `assisted' deposition-evaporation $\emptyset + A + 
\emptyset \rightleftharpoons A + A + A$,  as schematized respectively in 
Figs.\,\ref{Fkinks}a and \ref{Fkinks}b. The first situation represents the 
meandering spins referred to above and involves no energy changes (rates $1/2$). 
Notice that these dimers (double kinks) do not preserve their identity as they may
eventually dissociate when contacting a domain wall (single kink). The second 
process corresponds to a spin detachment (attachment) from a domain wall, and 
its rates $\epsilon_{_ K}$  ($\epsilon_{_ {\! - K}}$) parallel those of 
Eq.\,(\ref{rates}) when $s_{i-1} = - s_{i+2}$, thus
\begin{equation}
\epsilon_{_ {\!\pm K}} = \frac{1}{2}\,\left(\,1 \mp \tanh 2K\, \right)\,.
\end{equation}

Therefore, following the reasoning steps discussed for the spin representation 
and reinterpreting kinks as $\frac{1}{2}$-spinors in a dual chain, clearly the 
non-diagonal terms  $H^a_{nd} + H^b_{nd}$ of the evolution operator associated
to the (a) and (b) processes of Fig.\,\ref{Fkinks} can be constructed as
\begin{eqnarray}
\label{double-kink}
H^a_{nd} &=& -\frac{1}{2} \sum_i  \hat n_i \left(\, \sigma^+_{i+1}
 \sigma^-_{i-1} + {\rm h.c.} \right)\,,\\
\label{dep-eva}
H^b_{nd} &=& -\sum_i \hat n_i \left(\,\epsilon_{_K}\, \sigma^+_{i-1}
 \sigma^+_{i+1} + \, \epsilon_{_{\!-K}}\, \sigma^-_{i-1} \sigma^-_{i+1} \right)\,,
\end{eqnarray}
with number operators acting here as projectors that rule out vacancy mediated 
processes. To recast the non-diagonal operator into a symmetric representation we 
recur once more to detailed balance and rotate all $\sigma^{\pm}$\,'s around the 
$z$  direction using a  common pure imaginary angle $\varphi = i K$.  This is 
carried out by the non-unitary similarity transformation 
\begin{equation}
\label{i-rotation}
{\rm S} = e^{ \frac{K}{2}\,\sum_j \sigma^z_j}\,.
\end{equation}
for which it is straightforward to show that 
\begin{equation}
\sigma^{\pm}_{i-1} \sigma^{\pm}_{i+1}  \to 
e^{\pm 2 K} \sigma^{\pm}_{i-1} \sigma^{\pm}_{i+1}\,,
\end{equation}
thus producing the symmetrization of  $H^b_{nd}$ while keeping  all terms of 
$H^a_{nd}$ (already hermitian) unaltered.

As before, preservation of probability is taken into account by a diagonal operator 
$H^a_d + H^b_d$ balancing both of the above (mutually exclusive) events, in turn
probed respectively by
\begin{eqnarray}
H^a_d &=& \frac{1}{2} \sum_i \left[\,\hat n_{i-1}\, \hat n_i  \left( 1 - \hat n_{i+1} 
\right) \, +\,  \left( 1 - \hat n_{i-1}\right) \hat n_i \, \hat n_{i+1}\, \right]\,,\\
H^b_d &=& \sum_i \left[\,\epsilon_{_K} \left(1-\hat n_{i-1}\right)\, \hat n_i \, 
\left(1-\hat n_{i+1}\right)\,  + \, \epsilon_{_{\!-K}} \, \hat n_{i-1}\, \hat n_i \,
\hat n_{i+1} \, \right]\,.
\end{eqnarray}
Since the imaginary rotation (\ref{i-rotation}) is also diagonal, it has no effect
on these latter operators. Thus, collecting all terms and using PBC throughout, 
after some algebraic steps we finally obtain a self adjoint representation
for the kink evolution operator, namely
\begin{eqnarray}
\nonumber
{\rm H}_{kink} &=&  - \, \frac{1}{8} \sum_i \, \big(\,1+\sigma^z_i \,\big)\, 
\left[\, \left (\,1+ \gamma_{_K}\right) \,
\sigma^x_{i-1}\, \sigma^x_{i+1}  + \left (\,1- \gamma_{_K}\right) \, 
\sigma^y_{i-1}\, \sigma^y_{i+1}\, \right] \\
\label{hermitian-kink}
&+& \frac{1}{4} \sum_i  \left(\,1 + \alpha_{_K} \, \sigma^z_i\,  \sigma^z_{i+1}
 \right) + \frac{1}{2} \,\epsilon_{_{-K}} \sum_i \sigma^z_i \;\; , \;\; K > 0\,.
\end{eqnarray}
As can be readily verified from this equation, in the limit $T \to 0^+$ the action 
of ${\rm H}_{kink}$ on any configuration having non NN kinks vanishes like 
$O(e^{-4 K})$, thus yielding a metastable state. In the spin representation this 
corresponds to configurations with domains lengths larger than the lattice spacing,
which amounts to hindering spins to detach and so diffuse through. In turn, from 
Eq.\,(\ref{hermitian-spin}) it can be checked that this situation also yields
metastable states of ${\rm H}_{spin}$ so long as $J > 0$. Clearly, the number of 
these configurations grows exponentially with the  system size, and most of them
take over the asymptotic dynamics with long lifetimes $\propto \xi^2$. This marks 
an important difference with respect to the AF dynamics to be introduced briefly
in Sec. II B.

As an aside, we finally mention that by construction ${\rm H}_{kink}$ not only 
preserves the parity of kinks $e^{i \pi \sum_j \hat n_j}$ (being even for PBC), 
but in turn satisfies
\begin{equation}
\Big[\, {\rm H}_{kink}\,,\, \sum_j \sigma^z_j \,e^{i \pi \sum_{k \le j}\hat n_k }\,
\Big] = 0 \,,
\end{equation}
which simply expresses the conservation of the total spin magnetization in the 
original system. In practice, for the numerical evaluation of gaps we will just  build
up the adequate kink states from the corresponding spin ones.

%________________________________________
\subsection{Antikinks: $J < 0$}
%________________________________________
While most of the above ideas and procedures applies to the AF dynamics as well,
ultimately the relaxation to equilibrium becomes faster than in the F case.  By 
changing from a description based on kinks to AF domain walls, i.e. to antikinks, we
can easily construct an evolution operator that for large times and LT 
regimes essentially describes a diluted antikink system.  As before, if we think of
antikinks as hard core particles $A$ (former vacancies $\emptyset$ under F
couplings), the basic processes now involve (i) next NN hopping, i.e. $A + \emptyset
+ \emptyset \rightleftharpoons  \emptyset + \emptyset + A$, and (ii) deposition -
evaporation `assisted'  by vacancies, that is $\emptyset + \emptyset + \emptyset
\rightleftharpoons A + \emptyset + A$. 

%_______________ Fig. 3 (Antikinks: schematic view) _____________________
\begin{figure}[htbp]
\vskip -8cm
\centering
 \includegraphics[width=0.65\textwidth]{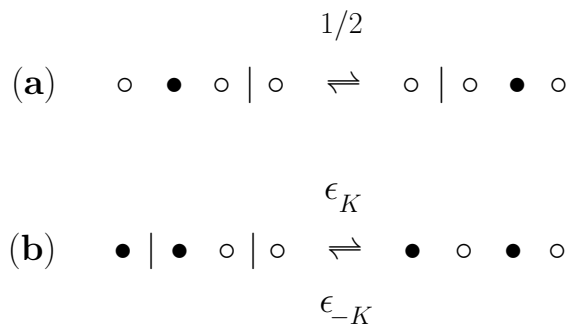}
\vskip -4cm
\caption{Transition rates for (a) next nearest neighbor antikink hoppings mediated
by vacancies, and (b) assisted  deposition-evaporation of antikinks. As in Fig. 2, 
these latter are represented by vertical lines but separating antiferro domains.}
\label{AFkinks}
\end{figure}
%______________________________________________________________________

With the aid of these schematic events, illustrated respectively in 
Figs.\,\ref{AFkinks}a and \ref{AFkinks}b, and after carrying out the rotation 
referred to in Eq.\,(\ref{i-rotation}) but with an argument $\bar\varphi = -i K$, 
the symmetric representation of the antikink operator can be finally cast as
\begin{eqnarray}
\nonumber
{\rm H}_{anti} &=&   \frac{1}{8} \sum_i \,  \big(\,1-\sigma^z_i \,\big)\, 
\left[\, \left (\,1+ \gamma_{_K}\right) \,
\sigma^x_{i-1}\, \sigma^x_{i+1}  + \left (\,1- \gamma_{_K}\right) \, 
\sigma^y_{i-1}\, \sigma^y_{i+1}\, \right]  \\
\label{hermitian-antikink}
&+& \frac{1}{4} \sum_i  \left(\,1 + \alpha_{_K} \, \sigma^z_i \,  
\sigma^z_{i+1} \right) - \frac{1}{2} \,\epsilon_{_{-K}} \sum_i \sigma^z_i  
\;\; , \;\; K < 0\,.
\end{eqnarray}
Of course, by exchanging the role of particles and vacancies, i.e. using the canonical
transformation $\sigma^{\pm} \to \sigma^{\mp}$, one recovers the form of 
Eq.\,(\ref{hermitian-kink}), but this likelihood is only superficial. Note that for 
$K \to -\, \infty$ the effect of the uniform field $\epsilon_{_{-K}}$ vanishes like 
$O (e^{4K})$, so the {\it only} jammed state on which the action of the diagonal
terms cancels out at this order is the vacuum of antikinks. Thus, in the limit 
$T \to 0^+$ the dynamics is no longer dominated by metastable states (instead 
proliferating exponentially for $J > 0$). In this situation AF domains can grow 
unhindered, as single antikinks now detach and diffuse freely (Fig.\,\ref{AFkinks}), 
thus allowing the dynamics to run smoothly towards equilibrium. As a result the
dynamic exponent, whose numerical analysis we next turn to consider, becomes 
smaller (actually diffusive) than for F exchanges.

%_____________________________________________________
\section{Numerical Results}
%_____________________________________________________

We now investigate numerically the spectral gaps of the Kawasaki operators 
given in Eqs.\,(\ref{hermitian-spin}), (\ref{hermitian-kink}), and 
(\ref{hermitian-antikink}) along with their dynamic exponents in different 
situations. As a consistency check  first we verified that the (rotated) Boltzmann
distribution corresponds in fact to the steady state of those operators. This also 
served to start up the Lanczos recursion with a random state but chosen orthogonal
to that equilibrium configuration.  Thereafter we obtained the first excited 
eigenmodes of our symmetric ${\rm H}'s$ using periodic chains of up to 24 sites,
the main limitation for this being the exponential growth of the space 
dimensionalities.  Another restrictive issue is that below  $k_B T / \vert J \vert \sim$ 
0.1 - 0.2\, the Lanczos convergence slows down progressively because in most 
situations the spacing of low lying levels turns out to decay as $e^{-4 \vert K
\vert }$. Thus, in what follows we content ourselves with giving results above that
region where nonetheless clear saturated tendencies can be already obtained.

%________________________________
\subsection{$J > 0$}
%________________________________

\begin{figure}[htbp]
\vskip -3cm
\hskip -2.5cm 
\includegraphics[width=0.65\textwidth]{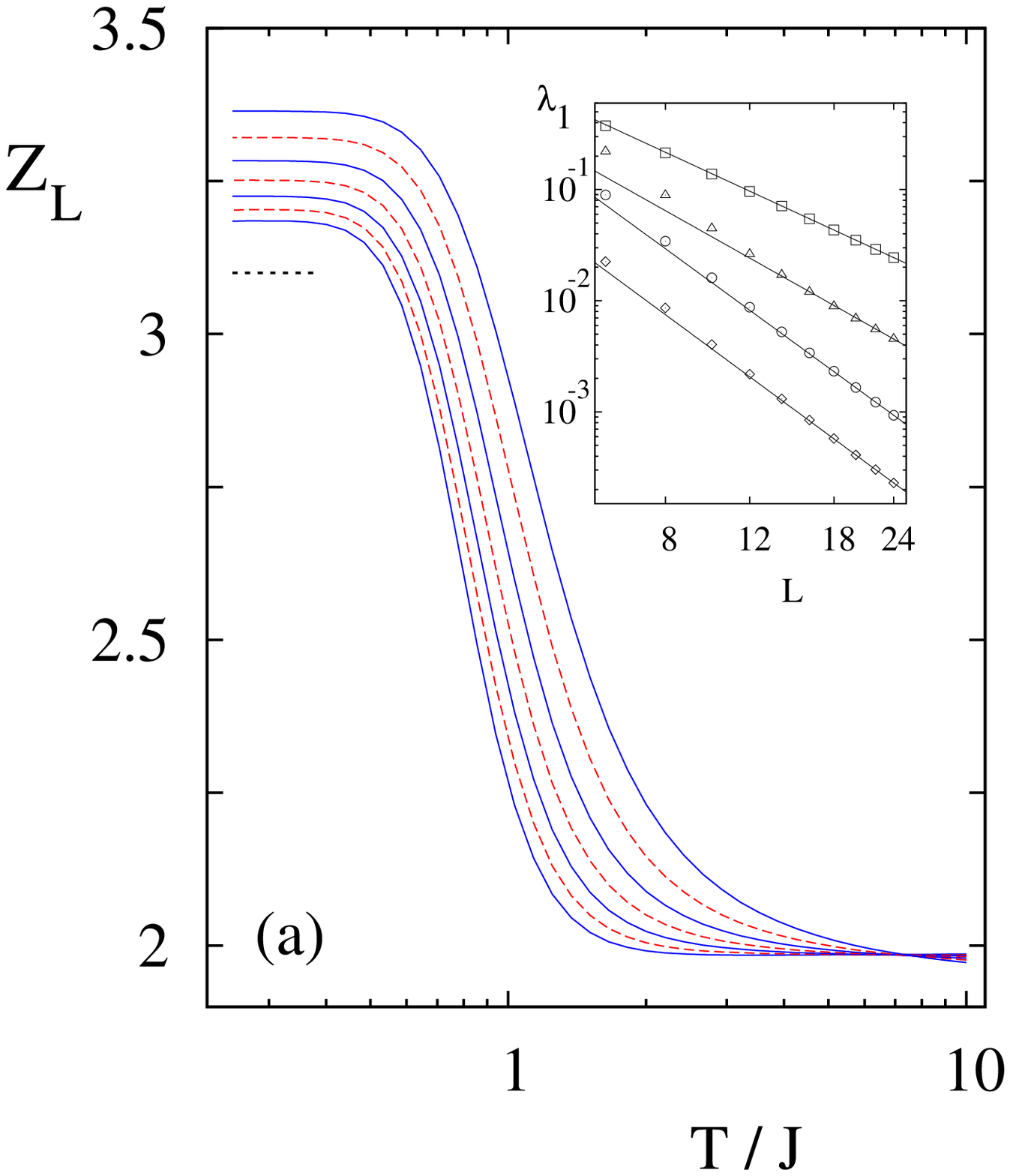}
\hskip -3cm
\includegraphics[width=0.65\textwidth]{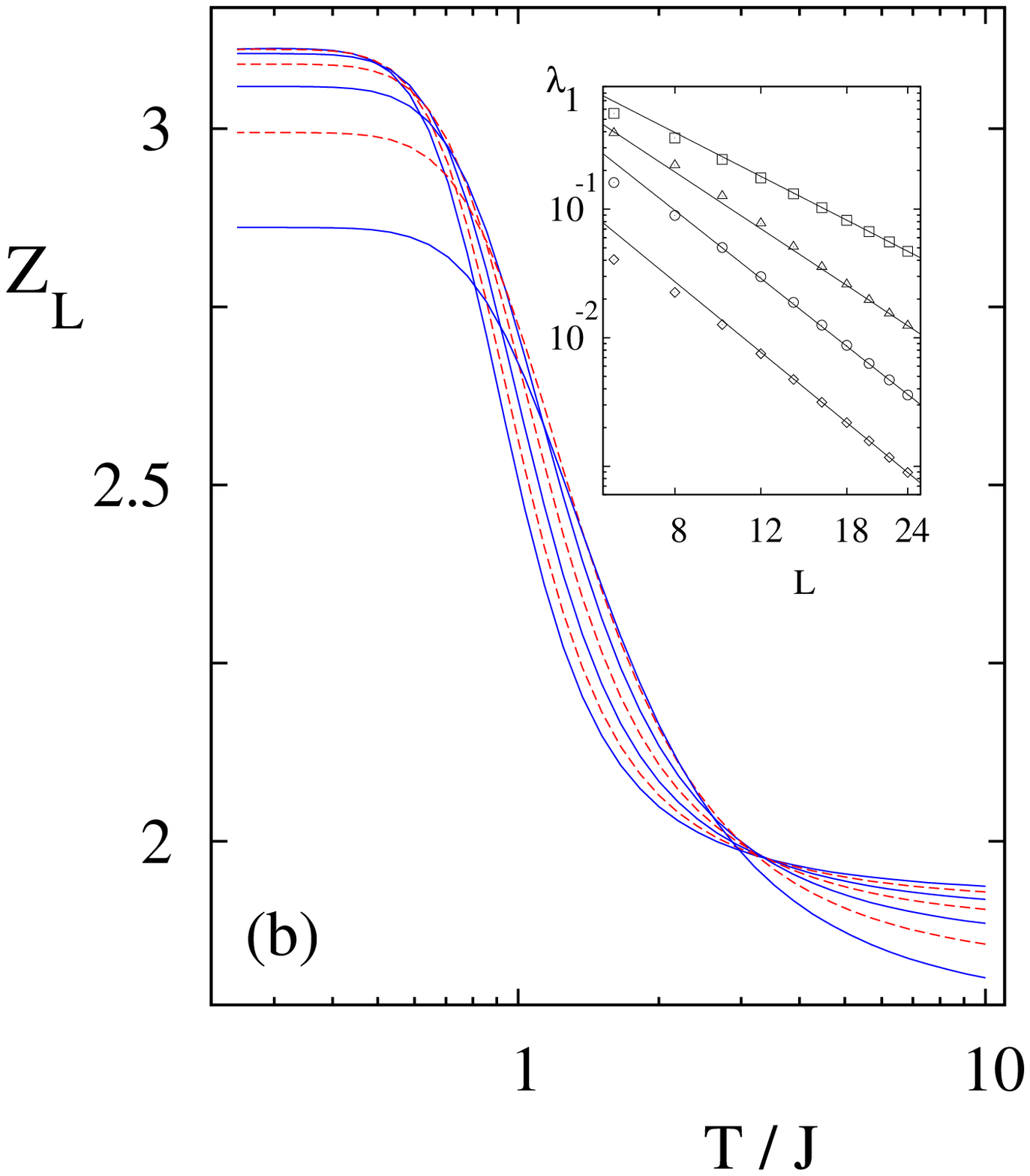}
\vskip -6cm
\hskip -2cm
\includegraphics[width=0.65\textwidth]{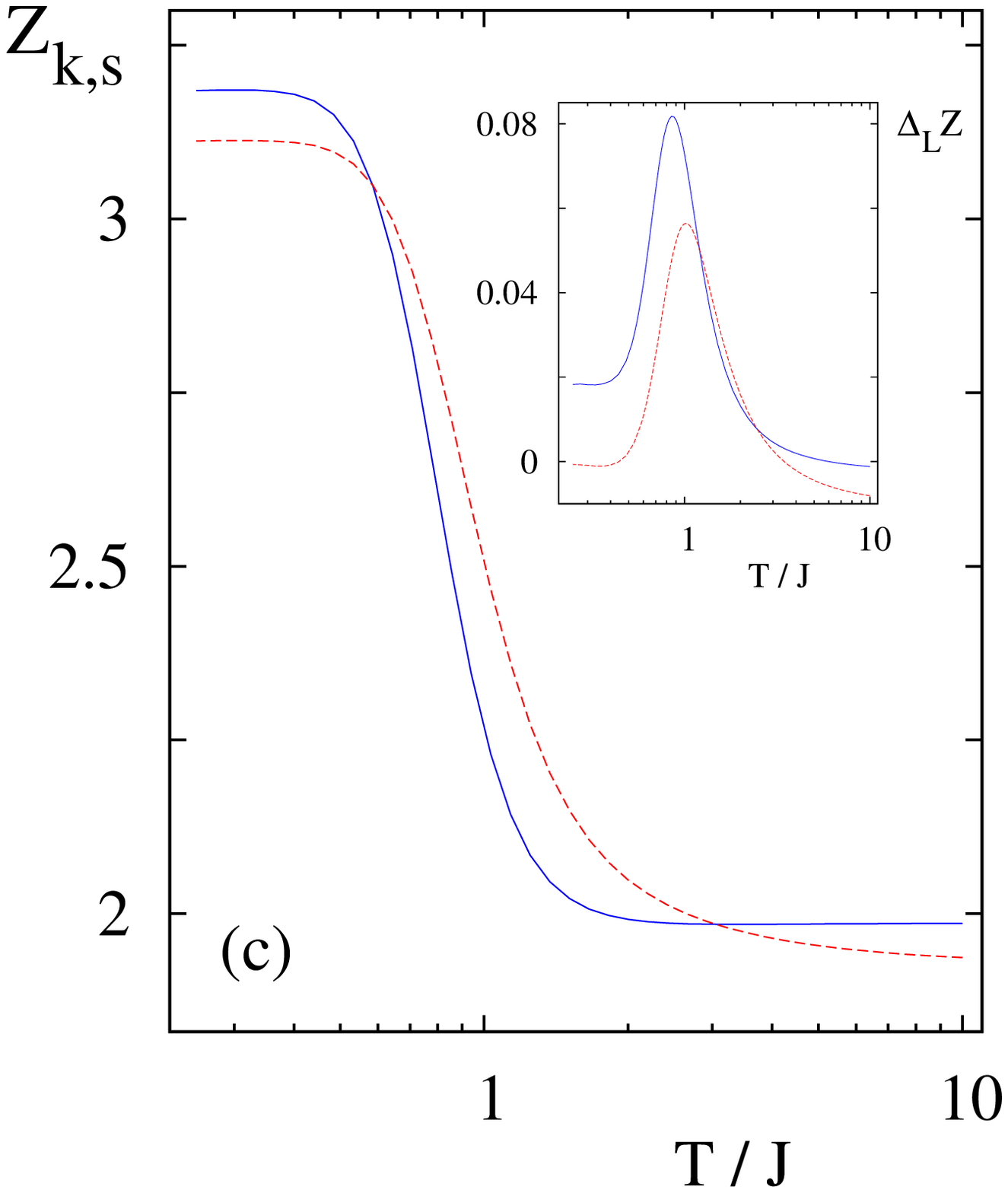}
\vskip -3.5cm
\caption{(Color online) Effective dynamic exponents of Eq.\,(\ref{approximants}) for
$J > 0$ using $12 \le L \le 24$ with $L$ even, computed for (a) spins, and (b) kinks.
Sizes increase  downwards almost  throughout (a), and upwards in (b) at both 
leftmost and rightmost  temperature regions. The respective insets display the gap
behavior with the system size for $T/ J = 10, 1, 1/2$ and  $1/4$ (top to bottom).
To compare slopes, data have been shifted upwards with respect to $T/ J = 10$. 
Straight lines are fitted with slopes  $-z_{24}(T)$ as calculated from the main
panels. In (c) we compare these exponents for kinks and spins (dashed and solid
lines). At low as well as at high temperatures $z_k$ and $z_s$ yield respectively 
lower and upper bounds for the thermodynamic limit of $z$. The inset of (c) 
provides a measure of convergence for these quantities by depicting $\Delta_L z 
\equiv z_{L-2} - z_L $ for our maximum sizes. Horizontal short doted line in (a)
stands for the extrapolations referred to in the text.}
\label{F-exponents}
\end{figure}

Turning to the F dynamics, in the insets of Figs.\,\ref{F-exponents}a and 
\ref{F-exponents}b we exhibit the finite-size behavior of spectral gaps for the case
of spins and kinks respectively. They are all consistent with a gap vanishing  like 
$1/L^z$, although with a nonuniversal temperature dependent dynamic exponent 
$z\,(T)$.  For the purpose of observing in more detail the trend of size effects on 
these exponents under different temperature regimes, we considered a sequence
of effective approximants defined as
\begin{equation}
\label{approximants}
z_L = \frac{\ln\left[ \,\lambda_1 (L-2) / \lambda_1 (L) \,\right]}
{\ln \left[ \,L / (L-2) \,\right] }\,,
\end{equation}
($L$ even), which simply provide successive measures of the gap closing in either 
of the above representations. These are shown in the main panels of
Figs.\,\ref{F-exponents}a and \ref{F-exponents}b. As expected, in the HT
region ${\rm H}_{spin}$ is able to yield fairly convergent estimations
of $z\, (\,\sim 1.99)$, so recapturing the plain diffusive limit above  $T / J \simeq 3$
(henceforth the Boltzmann constant $k_B$ is set equal to 1). Besides, in this 
representation the above approximants turn out to constitute upper bounds
of this exponent for almost all temperatures studied, though exhibiting
different spreadings. We attempted to extrapolate these bounds using several
procedures, but owing to the variable spreading this resulted in a rather noisy 
limiting curve. Nevertheless, in the region that most interests us the spreading
becomes more stable, thus upon using a van den Broeck - Schwartz extrapolation
scheme \cite{VBS} we found that  $z \sim \, 3.1(0)$, in reasonable agreement with
analysis and Monte Carlo simulations under instantaneous quenches \cite{Robin}.
Moreover, this result is also consistent with estimations of $z$ arising from
${\rm H}_{kink}$ approximants around the same region ($T / J \alt 0.4$) where,
as discussed above, the kink description in finite sizes is most reliable. In fact, no
extrapolations are needed here since already the values of $z_{20} \alt z_{22}
\alt z_{24}$ follow very closely one another, collapsing near a value of $z \simeq
3.11$. This is the main result of this section.

As an added bonus, note that in the LT regime the kink approximants 
come out conforming a sequence of lower bounds which conveniently complement
the upper ones provided by the spin representation. This is illustrated in
Fig.\,\ref{F-exponents}c where for $ T / J \alt 0.4$ both kink and spin approximants
should actually enclose the value of $z$ in the limit $L \to \infty$ (presumably much
closer to the kink border). Using the results of our maximum reachable sizes, this 
suggests that in the ordering regime the dynamic exponent is bounded as $3.11
\alt  z < 3.18$. For $0.5 \alt T / J \alt 2$ the trend of finite sizes of the kink 
approximants now reverses and, alike the spin  $z_L$'s, turn out to yield upper 
bounds of $z$. The kink approximants converge slightly faster in this region (see
inset of Fig.\,\ref{F-exponents}c), until for $T/J \agt 2$ the trend of bounds reverses
once more. Hence, at HT scales the actual value of $z$ is enclosed again by our kink
and spin approximants though this time these latter take the lead, as it should.

%__________ Fig. 4a, 4b and 4c (Z's for spins and kinks) ______ 
%_________________________________________________________________

\begin{figure}[htbp]
\centering
\vskip -0.5cm
\includegraphics[width=0.472\textwidth]{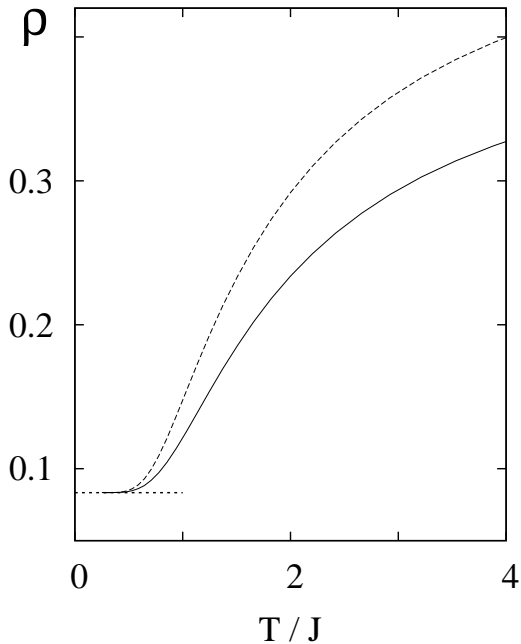}
\caption{Density of kinks for  the first excitation (solid line), and ground state
(dashed line) of ${\rm H}_{kink}$ [\,Eq.\,(\ref{hermitian-kink})\,] with $L = 24$ and
$J > 0$. Lowermost dots indicate the equilibrium density $(2/L$) when $T \to 0$.}
\label{kink-density}
\end{figure}

As mentioned earlier, the LT convergence of kink approximants was somehow
foreseeable on the basis of equilibrium considerations but it is not yet clear whether
these apply to nonequilibrium as well. In Fig.\,\ref{kink-density} however we show
that this also the case, at least for the sizes at hand. There, we display the kink
density $\rho_1 = \frac{1}{L} \sum_i \langle \psi_1 \vert \hat n_i \vert \psi_1
\rangle$ for the first excited mode of ${\rm H}_{kink}$ which for $T/J \alt 1$ closely
follows the diluted kink picture already expected for equilibrium, i.e. $\rho_1 \to 2/L$.
Also, preliminary evaluations of kink-kink correlations $\langle \psi_1 \vert \hat n_i 
\hat n_j \vert \psi_1 \rangle$ show that these become negligible below $T/J \sim 0.5$,
thus resembling the fully segregated equilibrium state. In passing it is worth 
pointing out that while $\lambda_1$ is doubly degenerate, the correlations and 
$\rho_1$ remain indistinguishable in both of these eigenmodes.
%_____________________ Fig. 5 (kink density) _____________________
%______________________________________________________________

\vspace{-0.5cm}

%________________________________
\subsection{$J < 0$}
%________________________________

\begin{figure}[htbp]
\hskip -0.5cm 
\includegraphics[width=0.472\textwidth]{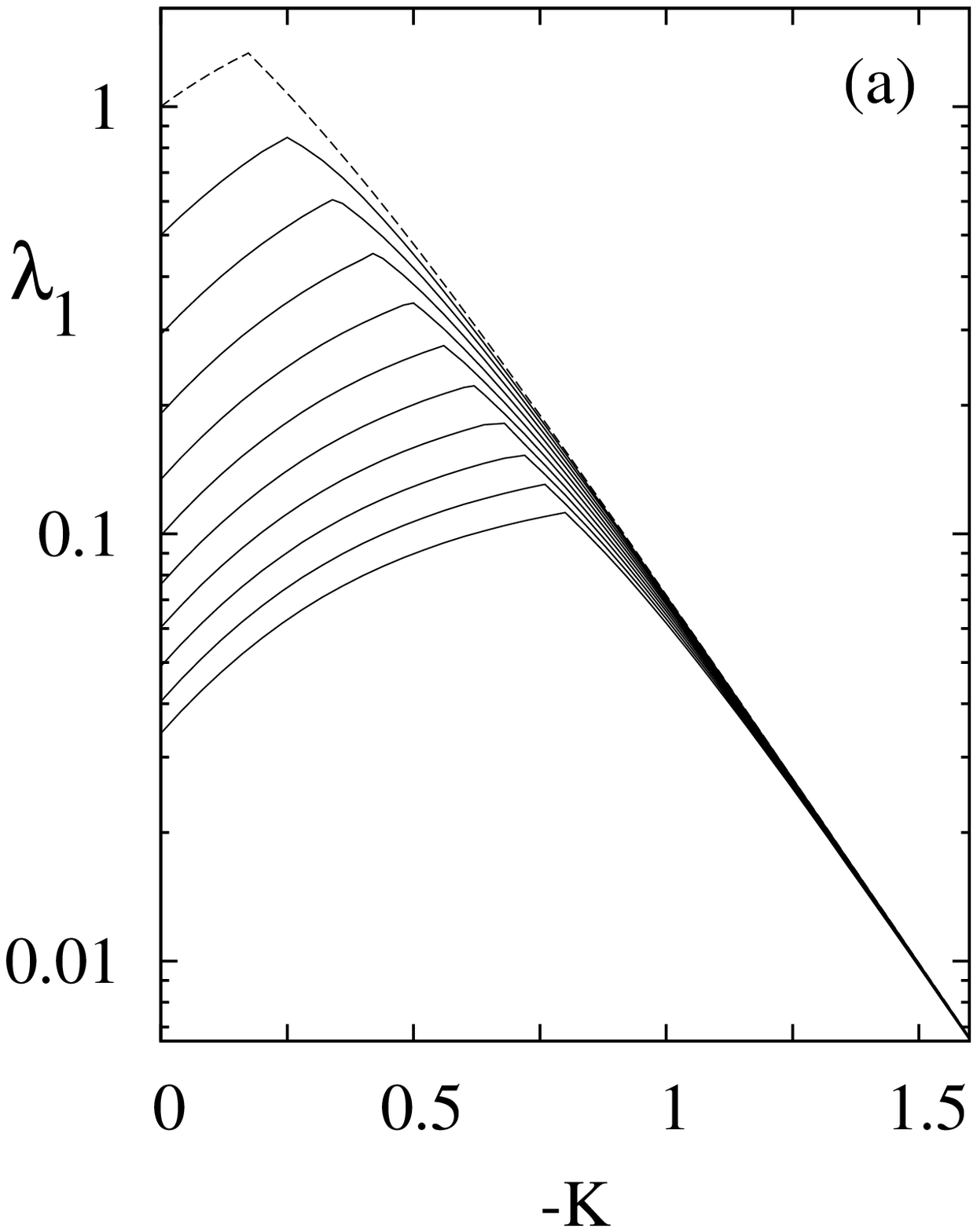}
\hskip 0.7cm
\includegraphics[width=0.472\textwidth]{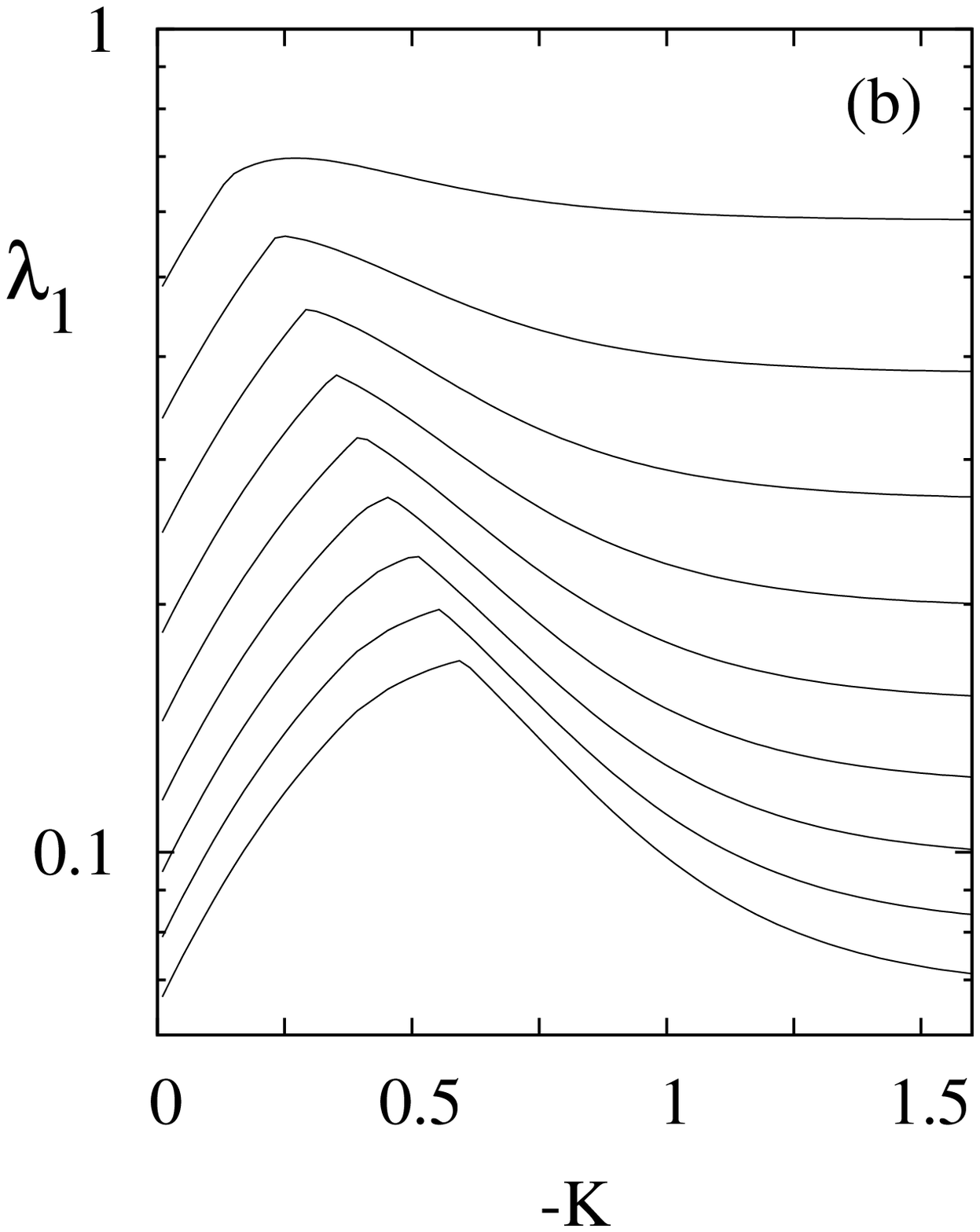}
\vskip 0.5cm
\caption{Spectrum gaps of (a) spin, and (b) antikink operators 
[\,Eqs.\,(\ref{hermitian-spin}) and (\ref{hermitian-antikink}) respectively\,], for 
various inverse temperatures $K = J/T$, ($J < 0$).  Sizes increase from top to 
bottom. In (a) the dashed line stands for the soluble case $L = 4$ already indicating
the appearance of a cusp due to level crossings [\,see Eq.\,(\ref{L=4})\,], a feature 
holding also for $L = 6,8,...,24$. At low temperatures $\lambda_1 (L)$ decays as 
$e^{- 4\vert K \vert}$ with an almost  size independent amplitude. In (b) this latter
problem is remedied although level crossings for $8,10, ... \,,$ and 24 sites now give
rise to non-analiticities near which size effects become severe
(cf. Fig.\,\ref{a-exponents}).}
\label{lambdas}
\end{figure}

In studying numerically the AF dynamics, the approximants (\ref{approximants})
pose severe difficulties at intermediate temperatures. Owing to low lying level
crossings, now the spectrum gap is bisected in temperature regions having opposite
monotonic behaviors. For instance, it is instructive to consider the case of $L=4$
where this feature already appears in the spectrum of ${\rm H}_{spin}$. By
diagonalizing its associated $6 \times 6$ stochastic matrix, we find that there
are two branches of eigenvalues crossing at $K^* = - \frac{1}{4} \ln 2$, thus 
producing the non-analytic gap 
\begin{equation}
\label{L=4}
\lambda_1 (4) = \cases{2\, \epsilon_K\;\;\,, \;{\rm threefold \;\;
                                   degenerate\;\; if}\;\; K \ge K^*\,,\cr 
                                    4 \, \epsilon_{-K}\,, \; {\rm nondegenerate\;\; 
                                    otherwise}\,,}
\end{equation}
as seen in the uppermost curve of Fig.\,\ref{lambdas}a. Alike this simple case, for
larger sizes it is found numerically that  $\lambda_1 (L)$ remains degenerate only 
above certain temperatures below which level crossings occur, and $\lambda_1$ 
results in a nondegenerate value. This is signaled by the emergence of the cusps 
observed in Fig.\,\ref{lambdas}a.  At LT regions the gap recovers the exponential
decay referred to above but the data collapse now precludes finite-size estimations
of effective exponents from Eq.\,(\ref{approximants}). Although level crossings
come out at successively smaller temperatures, these can not keep pace with the
increasing lattice sizes. To bypass these limitations we resorted to the antikink
operator especially constructed for this region [\,Eq.\,(\ref{hermitian-antikink})\,].
As can be seen in Fig.\,\ref{lambdas}b, now the LT spectrum no longer collapses so
the approximants (\ref{approximants}) can be employed once more. However, as 
evidenced by the cusps of that figure, the problem of low lying level crossings yet 
persists. In nearing the crossing temperatures this brings about pronounced size 
effects in $z_L$, partly because of the mismatches occurring between eigenvalue 
branches, alike those observed in Fig.\,\ref{lambdas}a. 

%_______ Fig. 6a and 6b (\lambda's for spins and antikinks )_____________
%_____________________________________________________________________

\begin{figure}[htbp]
\centering
%\vskip -7cm
\includegraphics[width=0.472\textwidth]{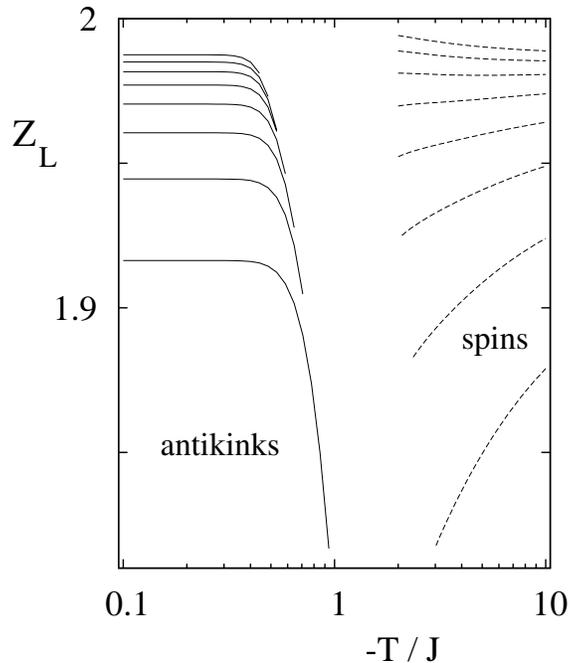}
\vskip 0.5cm
\caption{Effective dynamic exponents as defined in Eq.\,(\ref{approximants}),
for Hamiltonians (\ref{hermitian-spin}) and (\ref{hermitian-antikink}) 
with $J < 0$ and $L = 10, 12,\,...\,, 24$ increasing upwards. In nearing the 
cusps of Fig.\,\ref{lambdas} size effects become progressively pronounced, 
i.\,e. $\lambda_1 (L)$ and $\lambda_1 (L-2)$ belong to different branches. 
High and low temperature regimes are both consistent with a typical 
diffusive exponent $( z_{24} \simeq 1.99)$. }
\label{a-exponents}
\end{figure}

All the above considerations result in fair convergent estimations of dynamic 
exponents both for LT and HT regimes, but as displayed in 
Fig.\,\ref{a-exponents}, the intermediate zone is out of reach. Despite this setback,
what matters is the usefulness of the antikink representation 
(\ref{hermitian-antikink}) in the more interesting LT region which
otherwise would have remained inaccessible. As in the F case, no extrapolations to 
the thermodynamic limit are needed here since already for $T/\vert J \vert \alt 0.3$ 
our higher antikink approximants lie within the interval $1.98 \simeq z_{20} \alt
z_{22} \alt z_{24} \simeq 1.99$, thus indicating a Glauber universality class for 
this ordering regime, cf. \cite{Luscombe}.
%________________ Fig. 7 (Z's for spins and antikinks) ____________________
%__________________________________________________________________
In the other extreme, when disorder prevails, already for $T/\vert J \vert \agt 3$ 
the spin approximants derived from the data of Fig.\,\ref{lambdas}a converge
within one percent to our LT exponents, thus suggesting for these
latter a rather robust diffusive behavior. 

\begin{figure}[htbp]
\centering
\vskip -0.1cm
\includegraphics[width=0.472\textwidth]{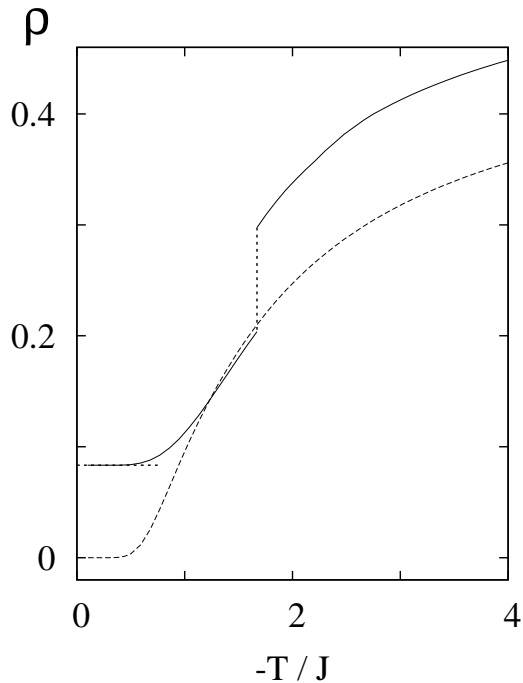}
%\vskip -3.5cm
\caption{Density of antikinks for the first level and ground state of  operator
(\ref{hermitian-antikink}) with $L = 24$ and $J < 0$ (solid and dashed lines
respectively). For comparison, horizontal dots indicate the equilibrium  density 
of ferro kinks $(J > 0)$ at $T=0$, cf. Fig.\,\ref{kink-density}. The discontinuity 
denoted by vertical dots stems from level crossings in the low lying spectrum of 
${\rm H}_{anti}$ (also, see cusp in the lowermost curve of Fig.\,\ref{lambdas}b).}
\label{antikink-density}
\end{figure}

Finally, in Fig.\,\ref{antikink-density} we exhibit the antikink density $\rho_1$ for
the first excited state of ${\rm H}_{anti}$. As before, this confirms the diluted 
antikink picture expected at LT scales, in turn accounting for the convergent
approximants obtained below $T/J  \alt 0.3$. However, this time $\rho_1$ does not
quite follow there the vanishing equilibrium density of the plain AF vacuum state
and adopts instead the limiting value referred to in Fig.\,\ref{kink-density}, 
indicating the presence of two antikink excitations.  At intermediate temperatures
the jump of $\rho_1$ just reflects the aforementioned level crossings.

%_______________ Fig. 8 (antikink density) ______________ 
%__________________________________________

\section{Concluding discussion}

To summarize,  we have constructed symmetric representations of the Kawasaki 
dynamics in the Ising chain using a quantum spin analogy with both direct an dual
processes. The kink description \cite{Robin}, either in its F or AF versions 
[\,Eqs.\,(\ref{hermitian-kink}) and (\ref{hermitian-antikink})\,], is well suited to low 
temperature regimes where it is able to provide a rather fast finite-size convergence
to dynamic exponents (Figs.\,\ref{F-exponents}b and \ref{a-exponents}). For F 
couplings, the ordering kinetics arising from these latter ($z \simeq 3.11$) tuns out
to be slightly slower than the Lifshitz - Slyozov behavior [\,$\xi (t) \propto t^{1/3}$]
characteristic  of higher dimensions \cite{Huse}. By contrast, for AF interactions
the kinetics is no longer activated by metastable states, and dynamic exponents 
converge rapidly ($z \simeq 1.99$) to the typical diffusive values of the Glauber
universality class [\,$\xi (t) \propto t^{1/2}$]. 

On the other hand, the spin representation [\,Eq.\,(\ref{hermitian-spin})\,] is more 
appropriate for higher temperature scales where the kinetics remains slow and is 
still characterized by relaxation times diverging as $L^z$ (inset of 
Fig.\,\ref{F-exponents}a). For the F dynamics, the finite-size approximants of these 
exponents turn out to form a sequence of upper bounds which also extends down to
low temperature regions (Fig.\,\ref{F-exponents}a), where it nicely complements the
series of lower bounds emerging from the kink approach (Fig.\,\ref{F-exponents}b).
On wider temperature scales, both representations are consistent with a 
nonuniversal set of dynamic exponents interpolating continuously between a 
subdiffusive ordering kinetics and the plain diffusive limit (Fig.\,\ref{F-exponents}c).
For $J < 0$, part of this intermediate region proved inaccessible due to size effects
caused by low lying level crossings in the spectrum of both spin and antikink
operators (cusps of Fig.\,\ref{lambdas}). However, as the dynamics of this case is
not activated, the diffusive exponents obtained by both representations within their
natural range of applicability (Fig.\,\ref{a-exponents}) suggest a rather universal
(Glauber) dynamics throughout.

The success of the kink description at low temperature regimes would probably
allow to extend the ideas of this work to other interesting $1d$-dynamics under 
instantaneous quenches (i.e. constant transition rates). For instance, the 
nonuniversal behavior alleged for the ordering kinetics of the alternating bond
Kawasaki chain \cite{Luscombe}, might well be further investigated with our
numerical approach.  Other lines of research that would also be worth pursuing are
certain dynamical processes involving composite objects (e.g. dimers) which exhibit
strongly broken ergodicity as a result of having an extensive number of conservation
laws \cite{Barma}. As the issue of universality classes in nonequilibrium statistical
systems is often linked to the existence of these latter \cite{Odor}, it would be 
important to determine whether dynamic exponents actually depend on the 
subspaces where the evolution takes place.  Dual representations constructed
appropriately for such processes could therefore provide reliable computations
of these exponents within ordering regimes. Further work in this direction is in 
progress.

%____________________________
\section*{Acknowledgments}
%____________________________

The author wishes to thank M. Arlego, G. L. Rossini, and F. A. Schaposnik for 
helpful comments and discussions. Support of CONICET, Argentina under Grants 
No. PIP 1691 and No. PICT ANCYPT 20350, is acknowledged.
%__________________________________________

%____________ References ___________________

\end{document}